# Risk Management for Distributed Arbitrage Systems: Integrating Artificial Intelligence

Akaash Vishal Hazarika[1*], Mahak Shah[2], Swapnil Patil[3], Pradyumna Shukla[3]

[1*]North Carolina State University, Raleigh, NC, USA.
[2]Columbia University, New York, NY, USA.
[3]IEEE Senior Member, Chantilly, USA.

*Corresponding author(s). E-mail(s): ahazari@alumni.ncsu.edu; Contributing authors: ms5914@caa.columbia.edu; swapnil.mobility@gmail.com; sppradyumna1@yahoo.com;

**Abstract**

Effective risk management solutions become absolutely crucial when financial markets embrace distributed technology and decentralized financing (DeFi). This study offers a thorough survey and comparative analysis of the integration of artificial intelligence (AI) in risk management for distributed arbitrage systems. We examine several modern caching techniques—namely in-memory caching, distributed caching, and proxy caching—and their functions in enhancing performance in decentralized settings. Through literature review we examine the utilization of AI techniques for alleviating risks related to market volatility, liquidity challenges, operational failures, regulatory compliance, and security threats. This comparison research evaluates various case studies from prominent DeFi technologies, emphasizing critical performance metrics like latency reduction, load balancing, and system resilience. Additionally, we examine the problems and trade-offs associated with these technologies, emphasizing their effects on consistency, scalability, and fault tolerance. By meticulously analyzing real-world applications, specifically centering on the Aave platform as our principal case study, we illustrate how the purposeful amalgamation of AI with contemporary caching methodologies has revolutionized risk management in distributed arbitrage systems.

# 1 Introduction

Considered as a natural aspect of financial markets, arbitrage trading is the exploita- tion of which traders may extract profit resulting from price differences among several trading platforms. Over these years, this development has been achieved with change in technology and different market conditions. Because of distributed systems, DeFi, and the integration of AI, new risks and opportunities have cropped up that raise questions about the validity of traditional ways of arbitrage.This is essentially an unpredictable nature of financial markets, in particular for cryptocurrencies, to be supplemented with good trading strategies, which offer less risk and higher returns. It becomes important to understand nuances in risk management tailored to the specifics of distributed arbitrage systems as traders get used to these new technologies.

# 2 Background and Related Work

## 2.1 The Inception of Conventional Arbitrage Systems

Traditional arbitrage in essence is the art and form , where traders make profit between asset pairs that have different prices in different regions. This has been studied quite a lot in literature with the base work laid out by Bachelier and Black publications. An important element present in those papers was the concept of risk while doing such exchanges. This is crucial because it would lead to lesser exchange losses. This was studied in detail by later academicians which expanded upon topics like market efficiency, hedging and transaction costs that impact the earning potential[1]. One such model that has a significant impact on traders' attitudes and behavior regarding risk is the Capital Asset Pricing Model (CAPM) [2].

## 2.2 Advancements in Modern Arbitrage Systems

High-frequency trading and the general adoption of computer algorithms have influenced the structure of arbitrage mechanisms. Some of the modern arbitrage strategies use complicated computational and predictive techniques to rapidly identify and exploit price discrepancies [3]. These changes facilitated an increase in the pool of investors who could benefit from arbitrage by enabling traders to respond swiftly to shifts in market conditions. On the other hand, the modern arbitrage systems being much faster and more complex also generated new problems such as flash outage and algorithmic errors [4]. Value-at-Risk (VaR) and Expected Shortfall (ES)[5] work hand in glove, symbiotically in modern distributed advancements.

## 2.3 Integration of Artificial Intelligence

AI and machine learning relevant to arbitrage trading have made the systems more robust. AI and machine learning are making it easier for arbitrage systems to quickly adapt to changes in the market. This helps them find complex trends in prices and find the best way to handle risk [6]. However, when AI and machine learning began to be integrated into arbitrage systems, several challenges and issues emerged, such as model overfitting, algorithmic failure risks, and other more efficient techniques of risk management [7]. This research contributes to the evolution of arbitrage systems by formulating and applying techniques of AI and machine learning in order to optimize risk management strategies in distributed arbitrage systems

## 2.4 Role of Smart Contracts

Smart contracts are central in the operation of decentralized exchanges, enabling automatic executions of transactions with no intermediaries. Smart contracts are pro- grammed to execute autonomously when predefined conditions are met; this improves the efficiency of transactions[8]. However, the complexity of the codes and inherent vulnerabilities pose critical operational risks.

AI can further empower smart contracts by embedding machine learning algorithms that dynamically change parameters on smart contracts due to market conditions for the best execution of trades with minimum attendant risks. For exam- ple, machine learning models analyze historical transaction data to predict the best conditions for executing particular trades. This can enable smart contracts to adjust dynamically in terms of execution strategy given current market conditions. Addi- tionally, AI increases smart contract security by providing active anomaly monitoring and informing data that can help developers improve code that may provide greater protection against such hacks.

Here's a simple example of pseudocode demonstrating how AI integration in a smart contract could work:

```
contract ArbitrageA I {
      DataAnalyzer dataAnalyzer ; // AI module for data analysis
      PriceFeed priceFeed ;  //Interface for real time price data

      function Arbitrage public(){
           // Analyze Market Conditions Using AI
           MarketTrends trends = dataAnalyzer.AnalyzeMarket();

           //Determine optimal thresholds for making trades
            uint OptimalBuyPrice = trends.getOptimalBuyPrice();

            uint OptimalSellPrice = trends.getOptimalSellPrice();

           // Fetch Current Market Conditions from feed data
           unit currentPrice = priceFeed.getCurrentPrice();

           // Only execute if  current price  will  yield  profit
           i f ( current Price >= optimalSellPrice ) {
                sellAsset( ) ;
           }
      }
}
```

Integration of AI into smart contracts, for operations such as real-time auditing and compliance checks, ensures trading operations are aligned with evolving regulatory requirements. Besides that, AI can be used to enhance the security of smart contracts by monitoring suspicious activities and creating actionable insights that help developers make the code in contracts resistant to potential exploits.

# 3 Risk Types in Distributed Arbitrage Systems

## 3.1 Market Risk

The financial success of an arbitrage operation is greatly influenced by changes in the asset prices across the exchanges. In highly volatile markets-a common feature with cryptocurrencies-market risk is huge. For example, when the price of Bitcoin skyrocketed in the second half of 2017, many traders tried to take advantage of the price gaps between various exchanges. However, due to extreme volatility, some trades were executed at very unfavorable prices, bringing huge losses for arbitrageurs who could not react on time. AI improves forecasting the shift in prices by analyzing histori- cal trends, sentiment from social networks, and news portals, thus enabling traders to make valued decisions. Sentiment analysis can also be used, for instance, to pre- dict sudden market fluctuations based on political events or some regulatory news. A striking example was when, in 2020, the market reactions for the listing of some alt- coin regulatory announcements drove their prices up and down.The traders employing AI-powered sentiment analysis picked up on trends, which signaled that increased regulatory scrutiny would be arriving earlier, allowing for the modification of arbitrage strategies in such a way that potential losses are minimized. Second, during the "Black Thursday" event that occurred in March 2020, when the cryptocurrency market[9] crashed overnight, most arbitrage traders took huge losses, as liquidity simply vanished in the space of a few minutes. AI algorithms tracking liquidity across exchanges could, therefore, warn traders against impending volatility and provide them an opportunity to hedge their positions. This again points to the important role that high-brow AI analytics now play, not only in making sense of the market risks but also in anticipating them, thereby making a case for the adoption of such technologies within distributed arbitrage systems.

## 3.2 Liquidity Risk

All issues with entry and exit of the market will obviously result in very serious losses, especially when there is distributed and, thus, potentially weak liquidity in the system. Of course, with pairs of less demanded assets, it would have a much higher bid-ask spread. For example, during periods of market turmoil-such as the Flash Crash in May 2010-some less liquid instruments rapidly crashed to a very low price and greatly cost the traders who did not have the time to liquidate their position. Machine learning algorithms can evaluate real-time market depth and hence help traders make decisions on entry and exit points with parameters on liquidity [10]. AI can analyze trading volume, historical liquidity patterns, and real-time market data to provide insights to traders on when to execute orders for minimal slippage and lower transaction costs. For example, during the volatility of the prices of altcoins in the boom of the ICOs in 2019, the traders who had AI-driven analytics were better equipped to handle the situation and hence could manage their liquidity risk by choosing the best time to trade. Also, AI can model alternative market scenarios and suggest the trade size to prevent the moving of prices. During the December 2017 rise in cryptocurrency price, traders who did not consider the risk associated with liquidity also placed large orders and saw huge changes in market price, which diminished returns as a result and increased costs of transactions. This can be done with AI simulation, whereby traders can strategize how to place orders in tranches to hold good price integrity, use large-scale arbitrage without igniting a liquidity crisis. Thus, the implication of machine learning toward liquidity risk management not only facilitates superior execution of trades but also controls the correlated risks arising due to inefficiencies of markets and surprise price movements.

### 3.3 Operational Risk

Failures in technology or process might heavily affect trading systems with considerable financial and reputational losses. This group includes but is not limited to risks of smart contract failures, network congestion, and system downtime that may impact the operations of trading. Consider the 2016 DAO breach, which resulted in the loss of approximately 60 million dollars worth of Ether and demonstrated how smart contract weaknesses may be exploited. This incident highlighted the importance of thorough testing and monitoring in decentralized applications. AI helps bring operational efficiency in finding system anomalies, faulty transactions, and slowdowns in the network quite early, thus enabling actions to avoid a big disruption.For instance, high-frequency trading systems may face great losses of money or opportunities when small latency or even minor errors occur. Machine learning algorithms can be used to allow trading organizations, by real-time monitoring of system performance, to identify sudden trends, such as spikes in latency or error rates, sooner and react faster to potential problems [11][12]. A good example was when, in May 2021, there was a huge outage with Binance Exchange. During that, no user could enter the site, and neither could any traders place their trade. It also caused disruptions not only among traders but also generally in the market, creating times of high volatility. AI Predictive Maintenance helps firms to figure out problems much before they have huge failures. Artificial intelligence can be incredibly useful to trading enterprises by allowing them to pre-schedule upgrades and maintenance using the historically available data, that would therefore allow less outage in peak trading periods.

### 3.4 Regulatory Risk

Changes in the regulatory framework can also bring huge uncertainty to arbitrage and general trading activity. The decentralized nature of many of the platforms puts them in a gray area of regulation, often making it very difficult to remain compliant. The 2017 ICO boom, for instance, saw the launch of many projects in murkier regulatory environments that eventually faced actions from regulatory bodies such as the SEC. The latter put many tokens under the category of unregistered securities and levied fines, thus shutting projects down. In navigating such complexities, AI can apply NLP[13] in real-time monitoring of updates on regulations, warning traders about compliance[14] matters and thereby minimizing potential legal problems. Companies like Chainalysis have begun to adopt AI-powered systems for analyzing and monitoring AML compliance by parsing legal documents and transaction history, keeping firms aligned with shifting regulatory standards. Apart from that, AI can generate comprehensive reports summarizing the requirements for compliance that will help an organization remain compliant with local and international regulations. A very good example of this is the automated reporting tools that assist a firm in compiling the necessary documentation for regulatory audits, hence making timely submissions and reducing the risk of penalties. It is especially relevant for environments like the Euro- pean Union, where GDPR imposes strict requirements on data treatment and privacy; the fines resulting from non-compliance can amount to millions.

### 3.5 Technical Risk

The heightened use of technology exposes distribution systems to numerous dan- gers, including cybersecurity threats, network attacks, and data integrity issues. Notable incidents, including the 2020 SolarWinds breach, have revealed weaknesses in significant organizations, resulting in illicit access to confidential data in essential government and financial sectors. Artificial intelligence-based cybersecurity solutions can turn out very much important in the reduction of these risks by helping them identify vulnerabilities and possible hazards. Active AI can investigate an enormous volume of network traffic to bring up peculiar trends that might imply a cyber attack. It includes DDoS attacks: Smashing systems with traffic. Using AI-driven cybersecurity systems helps companies identify trends suggestive of harmful activity, therefore enabling real-time threat responses. Companies like Darktrace apply artificial intelligence to create a "self-learning" system tailored to the specific actions of every user and device, producing instantaneous alarms when deviations happen. In one instance, a system from Darktrace spotted peculiar patterns of data access that did not con- form to the normal patterns of employees and thus rapidly identified and weeded out the insider threat. Artificial intelligence also enhances data integrity through predictive analytics and the detection of anomalies, providing advanced views into possible breaches before they happen. Through continuous monitoring and analysis of system logs, AI systems can detect anomalies or deviations from normal operating practices that indicate manipulation of data.

### 3.6 Reputation Risk

Of course, the most sensitive issues have to do with traders and platforms that are always much more in jeopardy from the aftermath of market collapses, security events, and major legislative actions. For instance, the infamous Mt. Gox 2014 hack resulted in the loss of some 850,000 Bitcoins and seriously damaged the overall reputation of cryptocurrency exchanges. This resulted in heightened regulatory requirements and a decline in consumer confidence. For quite a period of time following that incident, the cryptocurrencies entered the bear market. Reputation risk could reduce the sustain- ability of platforms and trading activity directly. Shrinking volumes of trade, increased regulatory scrutiny, and eventual withdrawal of liquidity coming with a tarnished rep- utation compromise viability in operating a platform. AI-powered sentiment analysis can be helpful in order to gauge the general opinion about either platforms in general or certain trading instruments. The algorithms are able to determine whether public opinion-through investigating online forums, news articles, and social media data-is positive or negative. A perfect example of such a platform is Brandwatch, which applies AI to provide companies in real time with information about consumer sentiment regarding their services. Moreover, through constant monitoring of social media and forums, a platform can easily contextualize a concern before it escalates into something far from trivial. For instance, if in a case of sentiment analysis there is a surge of negative discussions concerning one trading platform after the fall of the market, an organization can timely give a response by sending appropriate, clear messages on their stance and the actions that will be taken to change the course of events. It also provides insight into the up-and-coming trends or perceived problems of a trading instrument so that one may proactively change their strategy. An anticipa- tory reputation management strategy would do quite well for the brand of a platform and for its users in a market that keeps on changing at a breakneck pace.

## 4 Framework for Risk Management

This segment presents a detailed structure for handling risks in distributed arbitrage systems, particularly focusing on the integration of AI.

### 4.1 Risk Identification

Identification should therefore be a structured process of risk. The techniques include SWOT analysis, scenario assessment, and risk mapping, which give ample chance to the traders and organizations for identifying risks in advance. The hidden patterns and emerging risks buried under piles of big data volumes easily evade human analysts, which the application of machine learning techniques can only help excavate. This is further consolidated with the capture of crowdsourced feedback through social media analysis, which allows for real-time monitoring of market sentiment and trad- ing behavior through relevant tools, and much earlier warnings than when events may materially affect trading outcomes.

### 4.2 Methods of Risk Assessment

Quantitative AI-augmented models play an important role in assessing probability and impact in several risks. For instance, the application of neural networks to predictive analytics leads to enhanced metrics of risk like VaR and CVaR in turbulent cryptocurrency markets. AI enables an establishment of the correlation and causative elements to achieve deeper insight into exposure. It would then be easy to analyze the time-series data with reinforcement learning methods such that organizations, over time, adjust their risk models to accommodate new changes in the market scenario.

Given the very volatile nature of cryptocurrencies, model validation via stress testing and backtesting with historical data is highly essential. AI can automate such processes and do so more effectively by emulating various adverse conditions of markets that test performances under duress. This can reveal any underlying weaknesses in processes and approaches pertinent to the trading operation in issue.

## 4.3 Mitigation Strategies

The following may be identified as AI-supported strategies for successful mitigation of the identified risks:

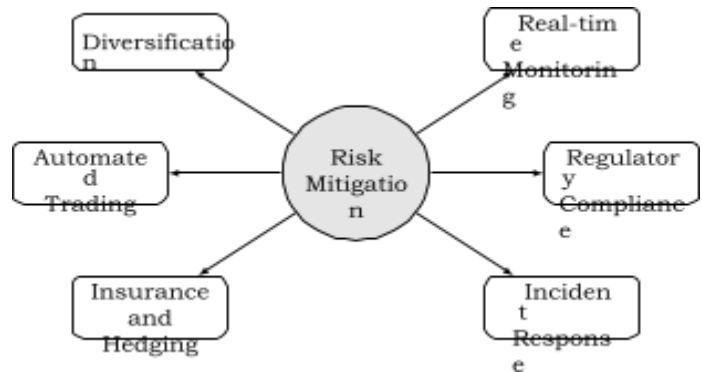

**Fig. 1** Core risk mitigation strategies in distributed arbitrage systems.

### 4.3.1 Diversification

While considering dynamically through AI the status of the market, an optimum diversification strategy thus could be suggested by distributing the investments among a variety of assets and trading platforms. This would bring down the exposure to a specific risk. It optimizes the weights over time by studying interlinkages between the assets.

### 4.3.2 Automated Trading Algorithms

AI-driven algorithms will be able to implement changes to the trading strategy in real time for any prevailing market condition and considerably reduce the possibility of mistakes by humans emotionally. The technology will automate the execution and optimization of trade placements with predictive analytics.

### 4.3.3 Insurance and Hedging Program

AI for market prediction helps identify better entry and exit timings of the hedging strategy, recognition of the most apt hedging instruments according to the market forecast. The capacity for prediction makes hedging economically viable and aligned with market conditions.

### 4.3.4 Real-time Monitoring Systems

AI-driven tools will instantly alert traders to market conditions, liquidity, and technical performance for prompt action when conditions are not so good. This could be in the form of analytics on a dashboard, a notification system displaying important information in real time, or AI triggering alerts when any of the thresholds set are exceeded.

### 4.3.5 Regulatory Compliance Framework

AI applied for adherence to regulatory standards will reduce risks of legal fines, trading halts, and operational halts. Automation of systems for constant regulation monitoring notifies them and gives recommendations of actions to be maintained. The integra- tion with the third-party regulatory knowledge base will enhance the accuracy and timeliness of compliance actions.

### 4.3.6 Incident Response Planning

AI-enabled incident response strategies will, therefore, be effective in helping the organization respond to any impending security breach or operational failure. Predictive analytics can help identify the common failure points where a firm could make investments in prevention. Often, the speed of decision-making reduces harm and preserves user confidence.

## 5 Infrastructure and System Design

The challenge of distributed arbitrage trading will only be properly answered by an AI-driven risk management system supported by a robust and scalable infrastructure. The system shall utilize several layers of comprehensive risk assessment and management.

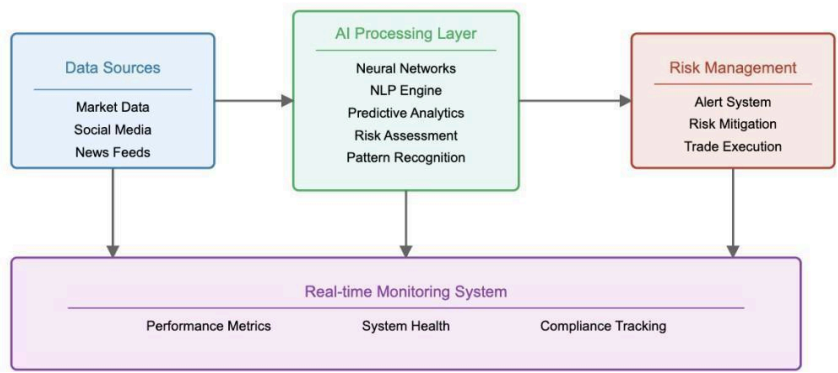

**Fig. 2** System Architecture for AI-Driven Risk Management - The four-layer architecture showing data input processing, AI analysis, risk management, and monitoring components.

## 5.1 Core Architectural Components

As shown in Figure 2, the system architecture comprises four fundamental layers that work in concert to deliver comprehensive risk management:

### 5.1.1 Data Input Layer

The data ingestion layer serves as the entry point for all market-related information. It includes things like social media analytics, market data feeds and historical data computation.

### 5.1.2 AI Processing Layer

This layer implements advanced machine learning capabilities. Some of the com- ponents include things like neural networks, NLP engine, pattern recognition and predictive analytics.

### 5.1.3 Risk Management Layer

This handles core risk operations. This include things like trade execution, risk mitigation and alert system.

### 5.1.4 Monitoring Layer

This section provides system-wide oversight. It includes things like performance metric, system health monitoring and compliance

## 5.2 Risk Assessment Data Flow

The risk assessment process follows a structured pipeline for data processing and analysis. Figure 3 illustrates this workflow:

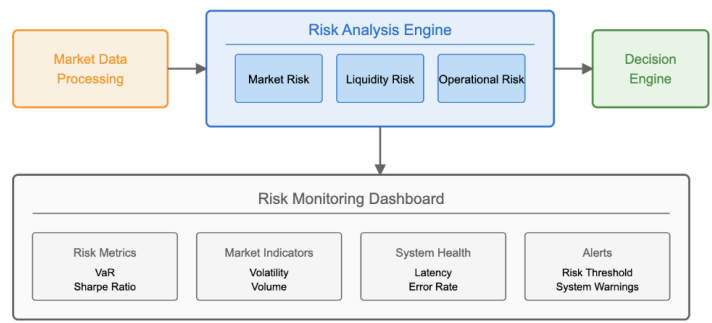

**Fig. 3** Risk Assessment Data Flow - Detailed visualization of the risk processing pipeline, from market data intake through monitoring and decision-making.

### 5.2.1 Market Data Processing

This section of the pipeline deals with stuff related to time series, data normalization, cleaning and anomaly detection. Initial data processing includes:

### 5.2.2 Risk Analysis Engine

The analysis engine computes operational, market and liquidity risks in real time.

### 5.2.3 Decision Engine

This section aggregates all information to find trading signal, combining all factors of computation. This also deals with adjusting some of the risk based params and trigger automated responses set beforehand.

### 5.2.4 Risk Monitoring Dashboard

This provides comprehensive visibility through having a 360 view over the system health and monitor the risk related metrics.

## 5.3 Implementation Considerations

The successful deployment of this infrastructure requires attention to several critical factors:

### 5.3.1 Performance Optimization

This deal with the optimization of the platform. Notable techniques include multi level caching [15][16][17], parallel processing and load balancing.

### 5.3.2 System Reliability Improvement

System reliability is improved via common approaches including redundancy, data replication and circuit breakers.

### 5.3.3 Security Framework

This deals with security frameworks related to audit system logging, data protection and some forms of access management.

# 6 Case Study: AI-Driven Risk Management in Aave

## 6.1 Implementation of AI Strategies

This case study examines the implementation of AI-enhanced risk management strate- gies in Aave decentralized market, a market based on the ethereum blockchain. This allows the market participants to lend and borrow cryptocurrencies.

- **Predictive Analytics:** Aave uses machine learning models studying historical lending and borrowing data to predict patterns in asset utilization and interest rates. This enables it to make far better decisions on the provision of liquidity, hence helping the platform optimize borrowing costs and offer maximum returns for lenders.
- **Liquidation Protection Mechanisms:** This involves Aave continuously monitoring the assets values in real time across the collateral securities. This would better result in risk containment due to fall in asset values resulting in margin-assignment.
- **Real-Time Risk Monitoring:** Aave's risk management infrastructure features a suite of AI-powered tools that analyze market trends and user behavior for the detection of any form of potential risks resulting from price volatility or market liquidity surprise. These monitoring systems create risk profiles around all user profiles, assuring high compliance and security.
- **Automated Compliance Systems:** This involves Aave in real-time monitoring through the govt websites and try to scrape in real-time related to any changes in the laws of a region. This indeed helps in the platform being compliant within a reasonable time period.

## 6.2 Results and Insights

The application of AI gave out remarkable results. Some of the major ones include:

- **Liquidation Events:** The platform under consideration: aave had a 30 percent decrease in non-essential liquidations. The above hypothesis can be tested by a two-sample proportion test comparing the liquidation rates during some period before versus after the AI went live. Further validation can be provided by a confidence interval for the difference between the two proportions.
- **Increased TVL:** By the start of 2023, Aave's TVL had crossed the $7 billion mark-huge growth in the amount of trust its users have put into its new and improved risk management features. This can be done as a time series analysis, comparing the mean TVL before and after the implementation of AI strategies, using a paired t-test to see the statistical significance. We expect to find a p-value less than 0.05. This will indeed give us strong evidence against the null hyptothesis.
- **User Growth:** The platform had grown quite a lot with around 1.6 million people dealing with lending and borrowing operations. A chi squared test would be suitable to showcase the strong dependence in growth of unique users based on leveraging an AI approach for risk management.
- **Increasing in Lending amount:** The total amount that were cumulatively borrowed by different users increased to $11 billion by mid-2021. The exact relationship between changes in the amount lent over time and deploying an AI strategy can be measured with a regression analysis, including confidence intervals for the coefficients to assess the precision of these estimates.

This case study illustrates how Aave's proactive approach to risk management, leveraging advanced AI technologies, can effectively mitigate risks in a rapidly evolving market while enhancing trading performance and operational security. The proposed statistical methods will allow for a more thorough validation of the impact of AI strategies, ensuring that Aave is well-positioned to navigate future challenges and opportunities.

# 7 Limitations and Drawbacks of AI in Risk Management

Although including artificial intelligence into risk management for distributed arbitrage systems has several advantages, it is important to underline its limits and possible negative effects for a fair view. The failure modes of artificial intelligence systems take front stage here. These models sometimes make improper assessments due to reasons such as overfitting in a model when an AI model is fitted too closely to training data, or failures in capturing the right subtlety of sudden market changes. The inaccuracies would lead to inappropriate risk measures and detrimental trading decisions.

Another major limitation would be the high computational cost associated with intensive AI algorithms. The materialization of advanced models requires high computational cost and time that could lower the capability of real-time decision-making, especially in high-frequency trading environments characterized by speed. In addition, AI-driven model interpretability is another challenge in the risk management process. A good number of AI techniques, mainly those grouped under deep learning methods, act like "black boxes" since stakeholders do not get to understand how such specific recommendations or decisions are arrived at.

There are several key limitations to consider:

## 7.1 Potential Failure Modes

Some of the common failure modes include model over fitting, where the model becomes closely tied to training data. This could also include failures related to adapting to market related volatility and changes.

## 7.2 Computational Overhead

There is a lot of computation overhead related to building such a platform. This includes and is not limited to high resource demands and the ability to make real time decisions.

## 7.3 Interpretability Challenge

Most of the AI systems behave sort of like a black box, where the key stakeholders of the project are unable to find a straightforward reasoning behind some of the decisions that an AI system had taken.

## 7.4 Reliance on Historical Data

Some of the historical data that these models have used to train might not contain the latest regulatory or volatility based changes. This makes the model harder to react when dealing with real time information for the first time.

In summary, while AI holds significant promise for enhancing risk management in distributed arbitrage systems, it is critical to remain aware of its limitations and challenges.

# 8 Conclusion

AI systems in the field of decentralized systems provide great value. The integration of AI competently into the processes of identification, assessment, and mitigation of risk will, therefore, enable traders to effectively handle the inherent challenges of such systems. Future research should be done on the development of overall risk metrics, considering all peculiarities of DeFi, enhancement of explainability in AI decision- making logic to establish stakeholder confidence, and building dynamic regulatory adaptation frameworks that can adjust themselves to dynamically changing compli- ance requirements. Further research will be likely performed in the field of AI-enabled cybersecurity measures that can be used to counteract threat actors. This will build resilience which in turn will transform the marketplace altogether.